\documentclass{WileyMSP-template}

\usepackage{hyperref}
\usepackage{url}
\RequirePackage{amssymb}

\hypersetup{
    colorlinks=true,
    linkcolor=blue,
    filecolor=magenta,      
    urlcolor=cyan,
    }

\begin{document}

\pagestyle{fancy}
\rhead{\includegraphics[width=2.5cm]{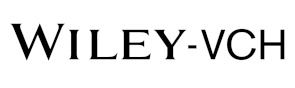}}

\title{Formation of amorphous carbon multi-walled nanotubes from random initial configurations}

\maketitle


\author{Chinonso Ugwumadu*,}
\author{Rajendra Thapa,}
\author{Yahya Al Majali,}
\author{Jason Trembly,}
\author{D. A. Drabold}


\dedication{}

\begin{affiliations}
C. Ugwumadu, Prof. D. A. Drabold\\
Department of Physics and Astronomy, \\
Nanoscale and Quantum Phenomena Institute (NQPI),\\
Ohio University, Athens, Ohio 45701, USA\\
E-mail: cu884120@ohio.edu\\

Prof. Y. Al Majali, Prof. J. Trembly\\
Russ College of Engineering and Technology, \\
Ohio University, Athens, Ohio 45701, USA

Dr. R. Thapa\\
Institute for Functional Materials and Devices\\
Lehigh University,Bethlehem, PA 18015, USA
\end{affiliations}



\begin{abstract}

Amorphous carbon nanotubes (a-CNT) with up to four walls and sizes ranging from 200 to 3200 atoms  have been simulated, starting from initial random configurations and using the Gaussian Approximation Potential [Phys. Rev. B 95, 094203 (2017)]. The important variables (like density, height, and diameter) required to successfully simulate a-CNTs, were predicted with a machine learning random forest technique. The models were validated \textit{ex post facto} using density functional codes. The a-CNT models ranged from 0.55 nm - 2 nm wide with an average inter-wall spacing of 0.31 nm. The topological defects  in a-CNTs were discussed and new defect configurations were observed. The electronic density of states and localization in these phases were discussed and delocalized electrons in the $\pi$ subspace were identified as an important factor for inter-layer cohesion. Spatial projection of the electronic conductivity favors axial transport along connecting hexagons, while non-hexagonal parts of the network either hinder or bifurcate the electronic transport. A vibrational density of states was calculated and is potentially an experimentally testable fingerprint of the material and the appearance of a low-frequency radial breathing mode was discussed. The thermal conductivity at 300 K was calculated using the Green-Kubo formula.

\end{abstract}

\section{Introduction}

  Owing to their unique structure, mechanical strength, variable electrical properties, and high thermal conductivity, carbon nanotubes (CNT) have been the focus of intense research with a range of applications in nano-scale engineering and electronics \cite{R1,R2,R3,R4_modelling,R5}. Methods of CNT production involving the catalytic decomposition of hydrocarbons in the presence of metal, as a by-product of arc-discharge or hydrocarbon flame production of fullerenes, laser vaporization, plasma-enhanced and thermal chemical vapor deposition, are well-established \cite{BAKER,OBERLIN,benito,iijima,pyrolysis,harrisbook}. The initial conception of CNTs as being perfectly seamless cylindrical structures of graphene layers without defects (except at the caps), has been shown to be oversimplified from experimental evidence \cite{EBBESEN}. Zhou and coworkers performed intercalation/deintercalation of potassium and rubidium on compressed CNTs and showed that the samples formed ``paper-mache" structures of small groups of graphite sheets instead of crystalline CNTs (c-CNTs) \cite{papermache}. Also, using electron nano-diffraction techniques, Liu and Cowley showed that both helical and non-helical CNTs exist, and the tube cross-section could be either circular or polygonal \cite{LIUPolygons}. In recent years, amorphous CNTs (a-CNT, CNTs with non-hexagonal rings that may not be represented by the Russian doll model \cite{russianJelly,russianJelly2}) have received increased attention due to their ease of growth \cite{aCNT_synthesis1,aCNT_synthesis2,aCNT_synthesis3} and their applications in areas like field-emission display devices, gaseous adsorbents, energy storage, and catalyst support materials \cite{aCNT1,aCNT2,aCNT3}. Evidently, the development of these a-CNT-based devices requires a good understanding of a-CNT’s properties at the atomistic level. This reason, coupled with the cost and technical difficulties associated with nano-scale experimental analysis, emphasizes the need to employ simulations as a tool for a preliminary and fundamental understanding of the atomistic structure and properties of a-CNTs. 
  
  An underlying motive for this work is discovering synthetic forms of graphite from available carbonaceous materials like coal. Carbon allotropes have a remarkable proclivity to form layers, whether as graphene layers in graphite, spherical shells in "buckyballs" fullerenes, or cylindrical tubes in CNTs. In earlier works, we predicted the formation of amorphous graphite, which is layers of graphene with 5- and 7- member rings in the hexagonal network \cite{LAG}, and multi-shell fullerenes with topological defects \cite{BO}. Theoretical studies of c-CNT are available \cite{RAFIEE2014435}, but this is not the case for a-CNTs. A notable exception is the work of Ghosh \cite{Ghosh}. In this paper, we explore the atomistic formation and properties of capped and hollow structures of Single-Wall a-CNT (a-SWCNT) and Multi-Walled a-CNT (a-MWCNT) obtained from initially random configurations of carbon atoms using the accurate density-functional-theory (DFT) trained machine-learning Gaussian Approximation Potential (\texttt{ML-GAP}) \cite{C} as implemented in the ``Large-scale Atomic/Molecular Massively Parallel Simulator" (\texttt{LAMMPS}) software package \cite{lammps}. The a-CNT models were validated using \textit{ab-initio} molecular dynamics (MD) packages including the plane-wave basis set DFT code, \texttt{VASP} (Vienna Ab initio Simulation Package) \cite{VASP} and the atomic orbital-based package, \texttt{SIESTA} (Spanish Initiative for Electronic Simulations with Thousands of Atoms) \cite{SIESTA}. The electronic structure and transport, as well as the origin of inter-wall cohesion in a-CNTs, are discussed in detail.  We explored vibrations in a-CNT by computing the density of states and their corresponding inverse participation ratio, phase quotient, and bond stretching character. Finally, we provide an estimate for the average thermal conductivity in a-CNT. We have also included some animations from our work and described them in sect.S1 in the supplementary material \cite{suppl}

\section{Computational Method}
\begin{figure}
	\centering
	\includegraphics[width=0.5\linewidth]{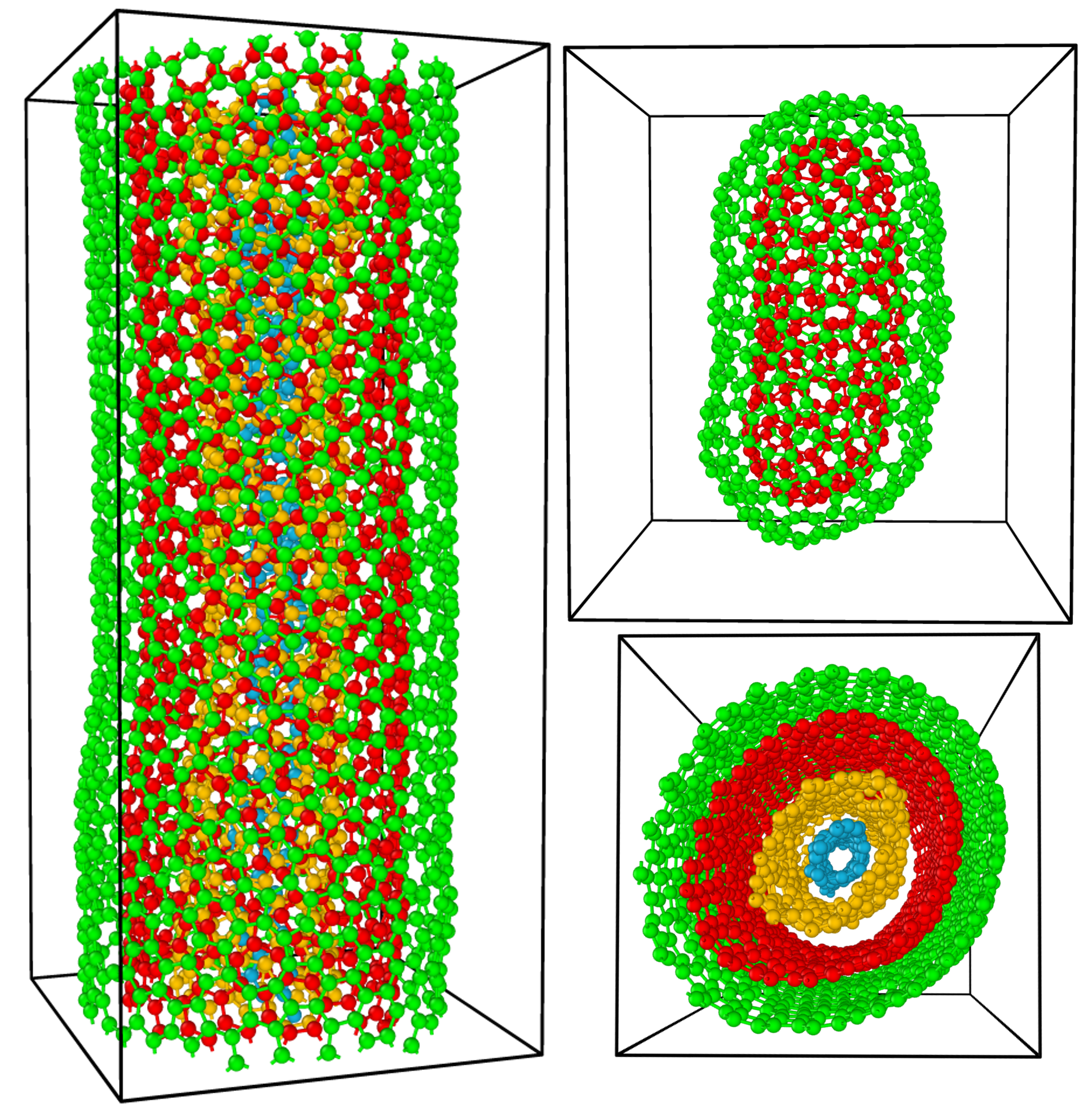}
    	\caption{Figure showing some a-CNT models with 3-fold coordination for each atom. The colors indicate the number of tubes in the a-MWCNT.}
	\label{fig:Cfig_nntubes}
\end{figure}

\begin{figure}
	\centering
  \includegraphics[width=.5\linewidth]{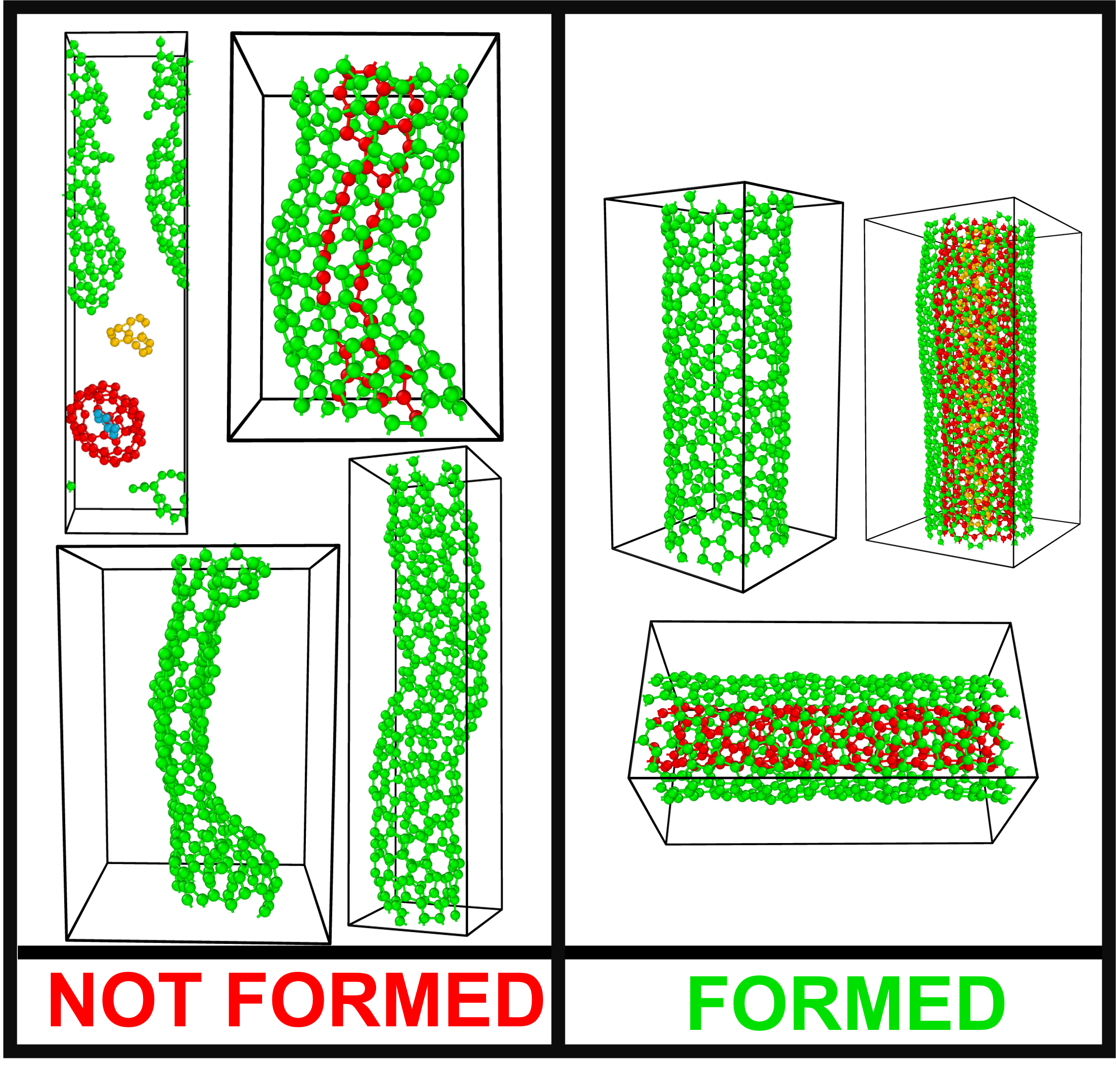}
  \caption{Examples of models that do not form [LEFT] and models that form [RIGHT] a-CNT based on their input variables}
  \label{fig:Cfig_examples}
\end{figure}

The \texttt{ML-GAP} C-C potential, which was trained on an extended set of reference data that was obtained from simulations based on first principle methods \cite{C,gap2,C2,C3,C4}, was used to simulate the amorphous carbon nanotubes. The starting models were arranged as a cylindrical bulk of randomly positioned carbon atoms at least 1.43 \AA~apart. The symmetry of CNTs required that the models were formed by including the periodic boundary condition (PBC) in one dimension, call it the $z$-axis, while the capped a-CNTS were formed by adding a narrow space of 3.5 \AA~in the $z$-axis between the cylinders. Without PBC in the $x$- and $y$- directions, the choice in the values of the input variables (like the density, diameter, height ...) required to successfully simulate a-CNT becomes non-trivial. To obtain good a-CNT models as shown in Fig. \ref{fig:Cfig_nntubes}, we trained a Random Forest (RF) classifier \cite{RF,randomforest,scikit-learn} on a set of forming/non-forming models to predict the important input variables (features) as well as their acceptable value limits. RF is a parallel ensemble learning  technique that combines multiple decision tree algorithms \cite{decisionTree} and involves a voting process that exploits those algorithms to obtain a better predictive performance. It uses random sampling of the training data set (with replacement) in building decision trees and for splitting the nodes in the network. We note here that the height (H) and diameter (d) of the models were substituted by a unitless parameter (aspect ratio; AR = H/d). At the initial stage, the parameters used in building the training dataset were sampled uniformly within reasonable limits (AR: 1-12; system size: 200 - 1500 atoms (in multiples of 40); vacuum in the $x$-$y$~plane: 1 - 10 \AA; cutoff radius: 1.4 - 1.7), which ensured that the initial configuration was constrained to form a cylindrical shape.  The final configurations obtained after MD simulations were labeled as "formed" (label = 1) or "not-formed" (label = 0) a-CNT. In addition to human validation, we automated the labeling process by using an "in-house" Model Filtering and Scoring Algorithm (MoFaSa), which assigns labels to models from a scoring metric. This metric is based on a combination of clustering analysis (to form non-connecting walls) and shape descriptors (like asphericity, acylindricity, and relative shape anisotropy) on a surface reconstruction of the atomic coordinates of each cluster using the alpha-shape method \cite{edelsbrunner} on a constructed Delaunay tessellation \cite{delaunay, stukowski} that estimate the shape of the model. An illustration of the decisions made by MoFaSa on different configurations is presented in Fig. \ref{fig:Cfig_examples}. With a total of 850 labeled models, the next step involved ranking the parameters on an importance scale to reduce the number of variables in building an additional training dataset. The feature importance is an implicit selection process implemented in random forest implementation that can be obtained using the Gini impurity criterion given as \cite{Gini}:

\begin{equation}
    \mathcal{G}(t) = 1 - \sum_{i = 1}^{m} \mathbf{P}(i|t)^2
\end{equation}

\begin{figure}
	\centering
  \includegraphics[width=0.7\linewidth]{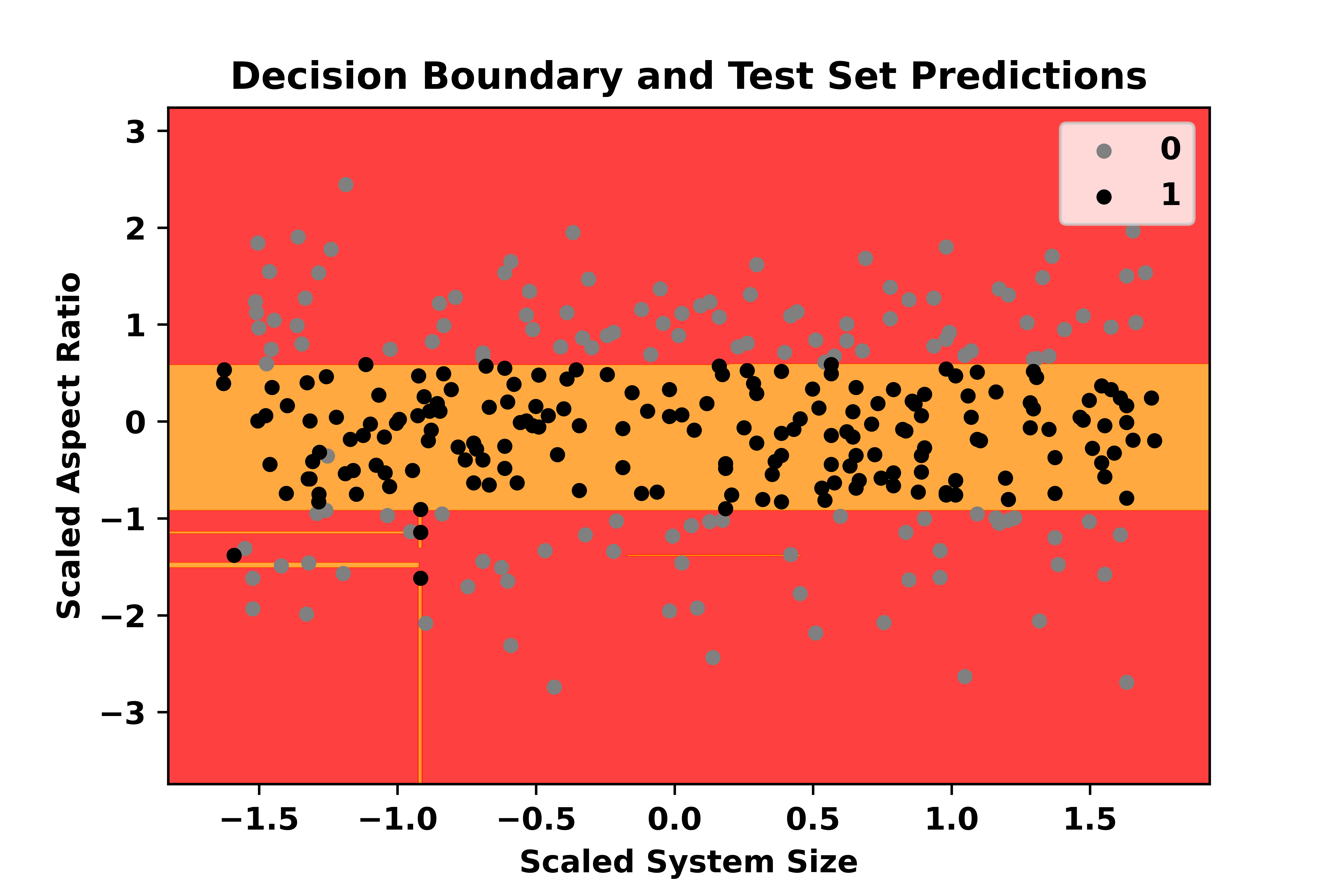}
  \caption{The decision boundary from a decision tree model using the aspect ratio and system size as features. 1 means an acceptable a-CNT formed, while 0 indicates otherwise. The feature values were scaled by subtracting the mean of the distribution and dividing the result by the standard deviation. The mean and standard deviation used for AR and system size are shown in Table \ref{tab:Ctable_RF}.}
  \label{fig:Cfig_DB_RF}
\end{figure}

\noindent The summation is over all the classes (t) in a training set m. The Gini scores for ranking the parameters (feature importance) for the K-fold cross-validation (K =4) set are shown in Table \ref{tab:Ctable_RF} and plotted as a radar chart in Fig. S1 in the supplementary material \cite{suppl}. The prediction revealed that the Aspect Ratio and system size are the most important features. We note that, while the average energy is obtained at the end of the simulation, it serves as an important feature during training since well-formed a-CNTs have lower total energies compared to more defective structures. Also, as our ML model is applied post-MD to find the feature importance, it is still reasonable to use the energy as a feature. Next, we trained a decision tree classifier to obtain a decision boundary that infers a range of suitable AR for the initial configuration (see a representative tree in Fig S2 \cite{suppl}. We generated 634 new training models with a different parameter distribution based on the previous results. The cutoff radius was fixed at 1.43 \AA. The amount of vacuum in the $x$-$y$~plane and the system size was uniformly sampled between 3 - 6 \AA~and 200 - 3200 atoms (in increments of 20) respectively, and AR was sampled in a normal distribution around a tested value of 4.3 as the mean and as a standard deviation of 1.2. An instance of the decision boundary is shown in Fig. \ref{fig:Cfig_DB_RF}. The orange region in the plot indicates that the optimal AR range is between 0.32 - 0.51 nm. The black and grey scatter points are predictions made by our ML model on a test set for a range of input variables that formed (value = 1) or did not form (value = 0) a-CNT. It shows that the decision boundary from our classifier makes a close to accurate prediction on unseen data. Before implementing the decision tree algorithm, the features were scaled by subtracting the mean of the distribution and dividing the result by the standard deviation. The mean and standard deviation used for AR and system size are shown in Table \ref{tab:Ctable_RF}. 
 
 For all a-CNT models in this work, the MD simulation protocol involved sampling the atomic positions and velocities on the canonical ensemble using a Nos\'e–Hoover thermostat at a fixed temperature of 3000 K for 240 ps with a timestep of 1 fs. Next, the models were allowed to find a more energetically and structurally favorable configuration by cooling to room temperature at a rate of 2.7$\times$10$^{13}$ K/s. We stress that the cooling rate does not alter bonds formed during/after heating. Finally, the structures were relaxed using a conjugate gradient (CG) as implemented within \texttt{LAMMPS} with a force tolerance of 10$^{-6}$ eV/\AA~to obtain a representative structure.  We also validated the energies and structure of the final models by independent CG relaxation using DFT-based Molecular dynamics (MD) packages, \texttt{VASP} and \texttt{LAMMPS}. The energy difference between the ML-GAP models and DFT-validated models ranged between 0.03 eV/atom $\leq \delta E \leq$ 0.07 eV/atom and DFT relaxation never led to bond breaking or forming, indicating that, in the configuration space, the atoms are formed in an energetically stable and realistic local environment. Model sizes ranged from 200 atoms to 3200 atoms, forming both a-SWCNT and a-MWCNTs (up to 4 walls). These systems will henceforth be referred to as a-CNT$_N$, where \textit{N} is the number of atoms. In the discussions that follows, we selected 10 independent models with N = 200 (a-SWCNT), 400, 840, and 2000 (a-MWCNT) atoms (40 models in total).

\section{Results and Discussion}
\subsection{\label{sec:structure}Formation and nano-Structure}
\begin{figure}
	\centering
	\includegraphics[width=0.7\linewidth]{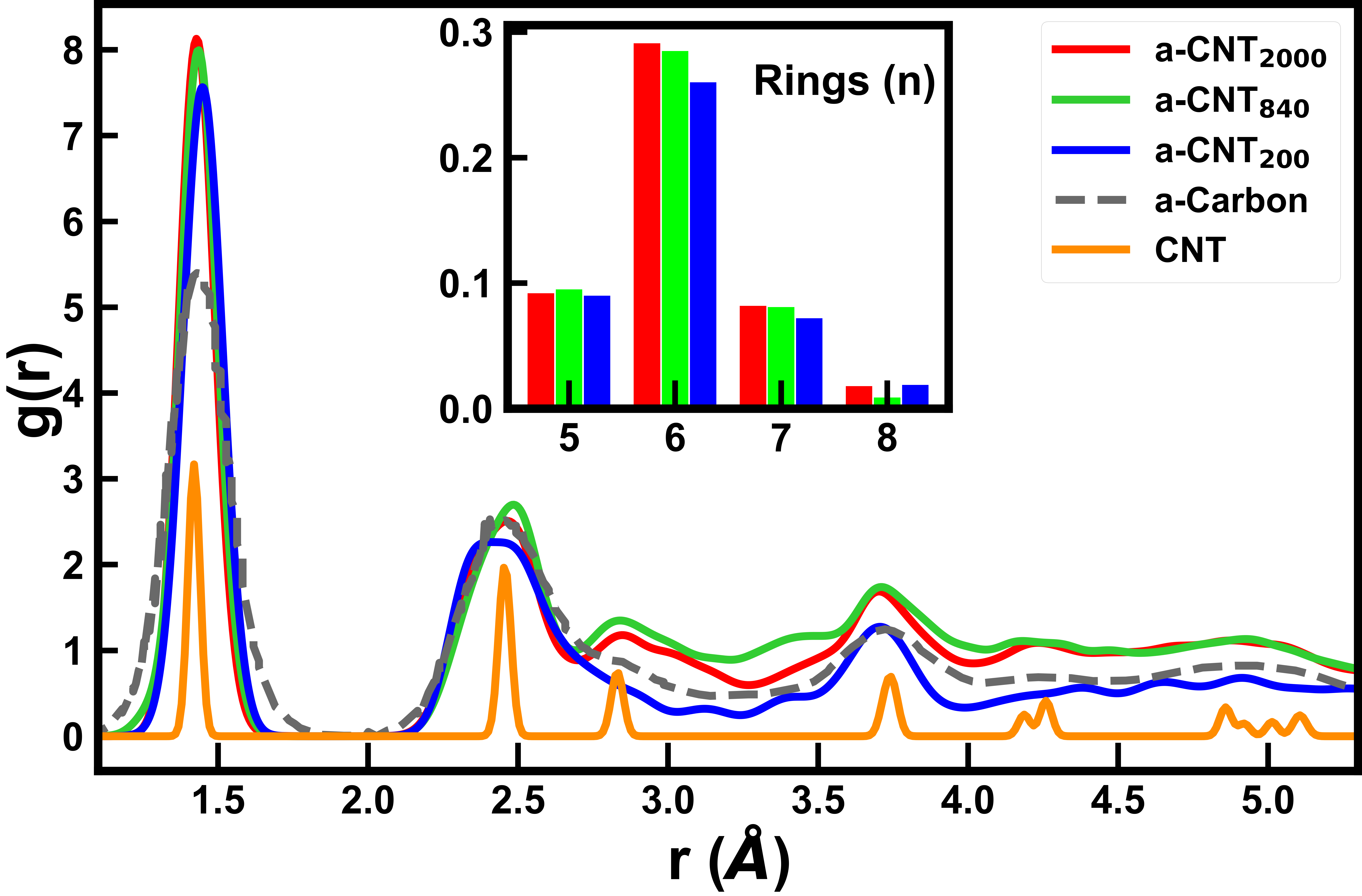}
    	\caption{Radial distribution function g(r) for different a-CNT models compared with crystalline CNT (c-CNT) and  amorphous Carbon (a-Carbon). The inset shows the ring distribution obtained using King's criterion for ring statistics.}
	\label{fig:Cfig_gr}
\end{figure}

We have provided animations showing the time evolution in the formation of capped and hollow a-CNT in the supplementary material \cite{suppl}. The a-CNTs were observed to form an outermost tube first, followed by inner tube(s). This formation process is similar to multi-shell fullerenes \cite{BO}. The final models have diameters ranging from 0.5 nm to 2 nm for a-CNT with up to 4 walls, and the average gallery (space between any 2 consecutive layers) in the a-MWCNTs models was calculated to be 0.31 nm, which is within the range that had been reported in experiments \cite{interwallSpacing} and close to the inter-layer spacing in graphite. The structural order of the a-CNT models was analyzed from the radial distribution function (RDF). In Fig. \ref{fig:Cfig_gr}, the peaks obtained for the a-CNT models were compared with those from crystalline SWCNT (c-CNT; n,m = 10,10) \cite{c-CNT}  and amorphous Carbon (a-Carbon) \cite{DAD_aC}.  The RDF peak for the first shell in all a-CNTs was close to the C-C bond length observed in c-CNT and graphite. Notably, the RDF of a-CNT$_{200}$, which is a single-walled a-CNT, has similar features resembling a-carbon and only the multi-walled a-CNT models reproduced the third c-CNT peak around 2.84 \AA. This is a consequence of the higher ratio of hexagonal to non-hexagonal rings (6:n; n = 5 or 7) observed in a-CNTs which have larger system sizes. The size of the topological defects, which are the non-hexagonal arrangements of carbon atoms incorporated in the hexagonal network was calculated using King's criterion for ring statistics \cite{KING} (see inset in Fig. \ref{fig:Cfig_gr}) and it showed similar values for all the a-MWCNTs with a slight decrease observed for a-SWCNTs. 

\begin{figure}
	\centering
  \includegraphics[width=0.9\linewidth]{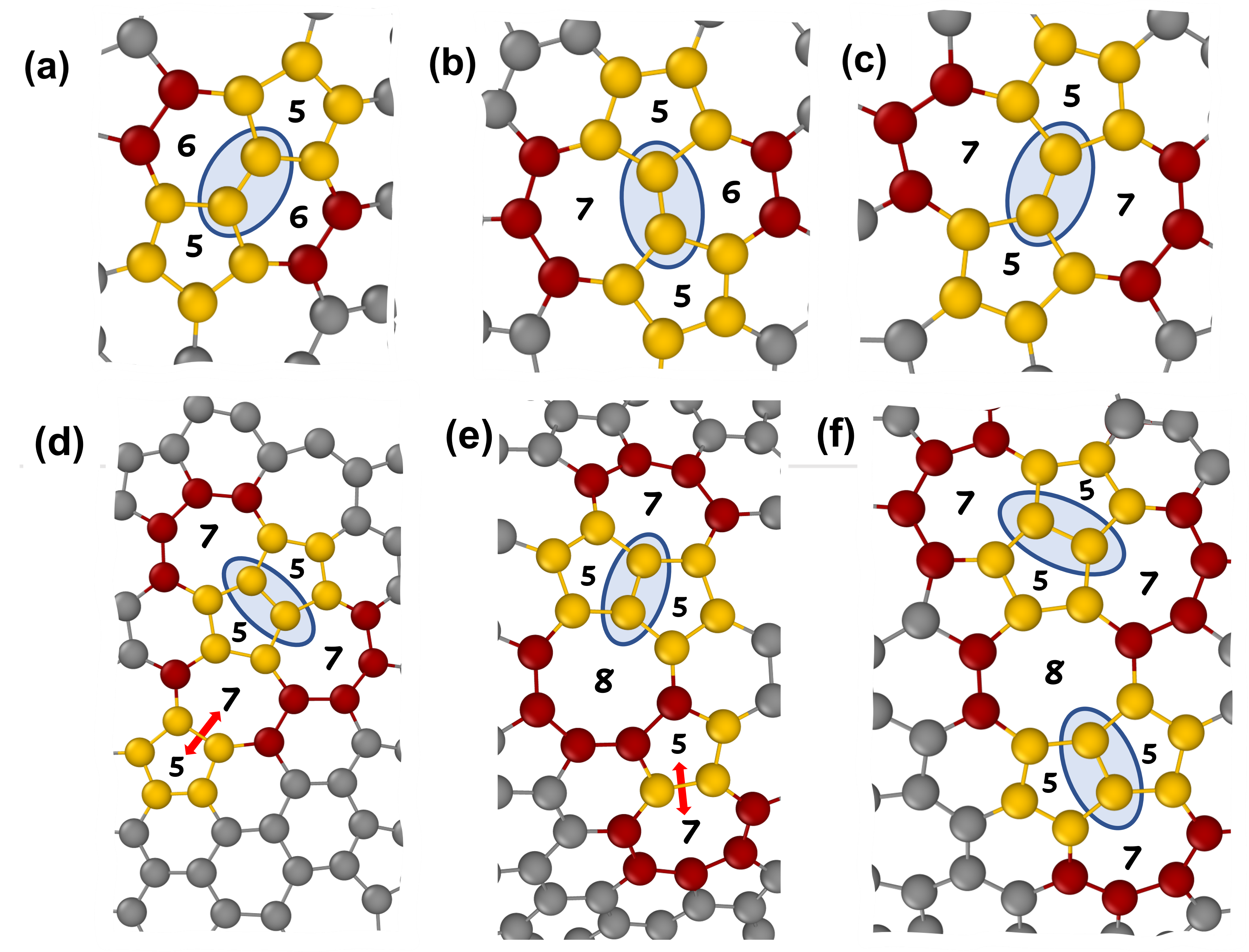}
  \caption{Figure showing primary ring defects in a-CNT. These defects are (a) Stone-Wales (5665) (b) 5675 and (c) Stone-Thrower-Wales (5775) defects, which are the primary defects. The secondary ring defects in a-CNT are the (d) inverse Stone-Thrower-Wales (7557) + 57 (e) 7558 + 57 (f) 7557 + 7558 defects. The blue ring indicates the connecting vertex atoms (a-c) or shared basal atoms (d-f) in pentagon-forming rings, from which we classify the defects.}
  \label{fig:Cfig_RINGDEFECTS}
\end{figure}

Defects in nanostructures like CNT play a role in their morphology and chemical properties. Experimentally and otherwise, an accurate and quantitative description of the defects or a standard for distinguishing them is still wanting. To this end, we studied the ring defects in the a-CNT models, Fig. \ref{fig:Cfig_RINGDEFECTS} (a-f) shows all the possible defects that exist in a-CNT. Fig \ref{fig:Cfig_RINGDEFECTS} (a), (b) and (c) are the Stone-Wales (SW; 5665), 5675, and Stone-Thrower-Wales (STW; 5775) defects respectively. We refer to these defects (a-c) as primary defects because they can exist alone in the hexagonal C network. Also, Fig. \ref{fig:Cfig_RINGDEFECTS} (d), (e), (f) are the inverse Stone-Thrower-Wales (ISTW; 7557 + 57) (e) 7558 + 57 (f) 7557 + 7558 defects. These are referred to as secondary defects since they are the union of two distinct defects. The blue ring indicates the two connecting vertex atoms (see Fig. \ref{fig:Cfig_RINGDEFECTS} a-c) or the two shared basal atoms (Fig. \ref{fig:Cfig_RINGDEFECTS} d-f) in pentagon-forming rings, from which we classify the defects. The 57, STW and ISTW defects are well-known defects in carbon nanotubes \cite{impurities,STW,STW2,ISTW,Thrower,Stone-Wales}, but to our knowledge, there has been no report on 5675 and 7558 defects, hence this is the first time such defects will be reported for any layered carbon network. We stress that the heptagon-pentagon and octagon-pentagon still preserve the connectivity of the sp$^2$ hybridized lattice and a challenge for future research is in the modification of these defects in MWCNTs to achieve specific morphologies with attractive electronic, optical, and thermal properties. 

\subsection{\label{sec:electronicstructure}Electronic structure, Transport and $\pi$-electron Delocalization}
\begin{figure}
	\centering
	\includegraphics[width=0.9\linewidth]{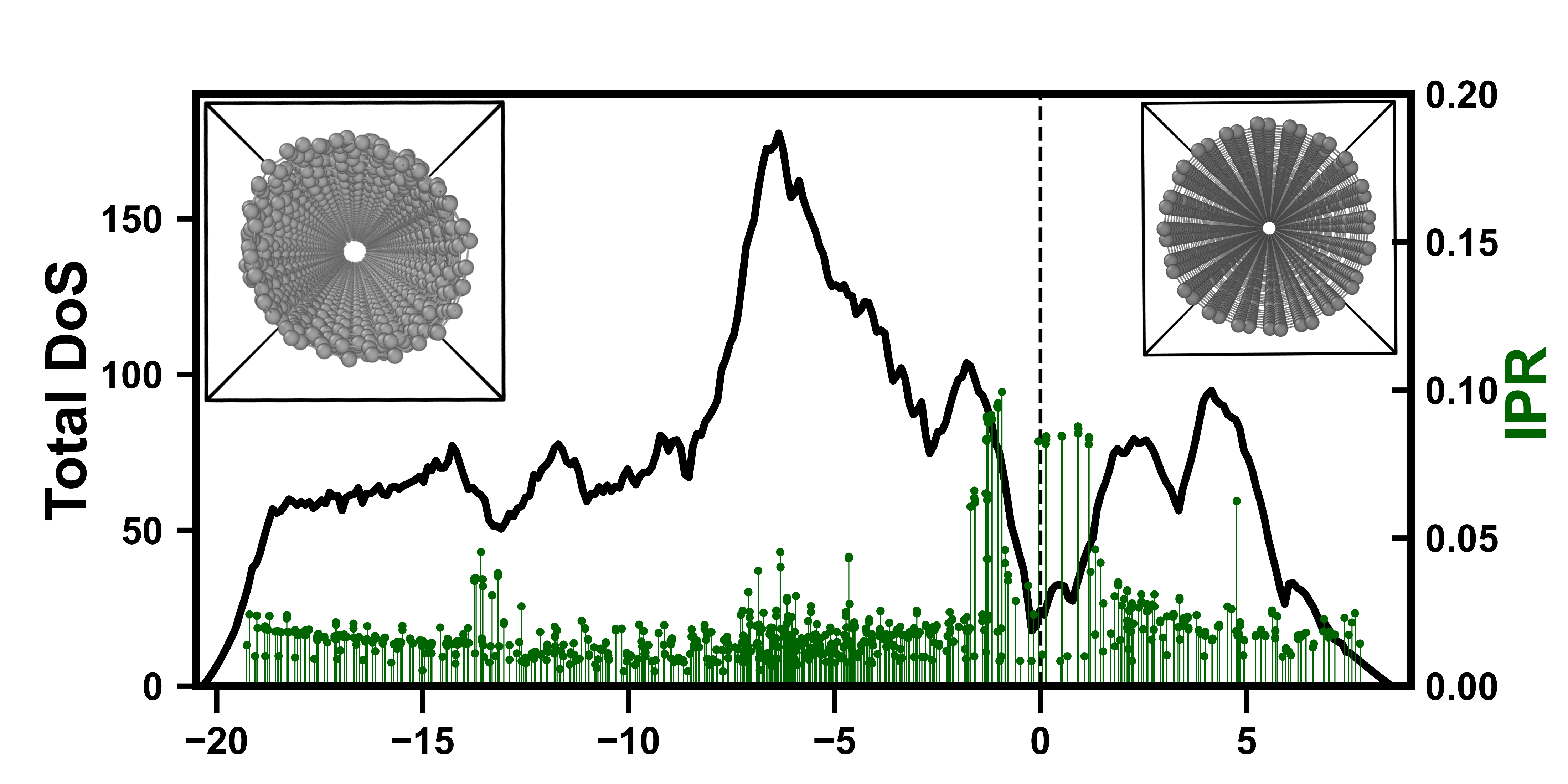}
    	\caption{Electronic density of states for a-CNT. The [LEFT] and [RIGHT] insets compare the structure of a-CNT and armchair CNT. The periodic images of the models were replicated in the $z$-direction.}
	\label{fig:Cfig_DoS_IPR}
\end{figure}

\begin{figure}
	\centering
	\includegraphics[width=0.5\linewidth]{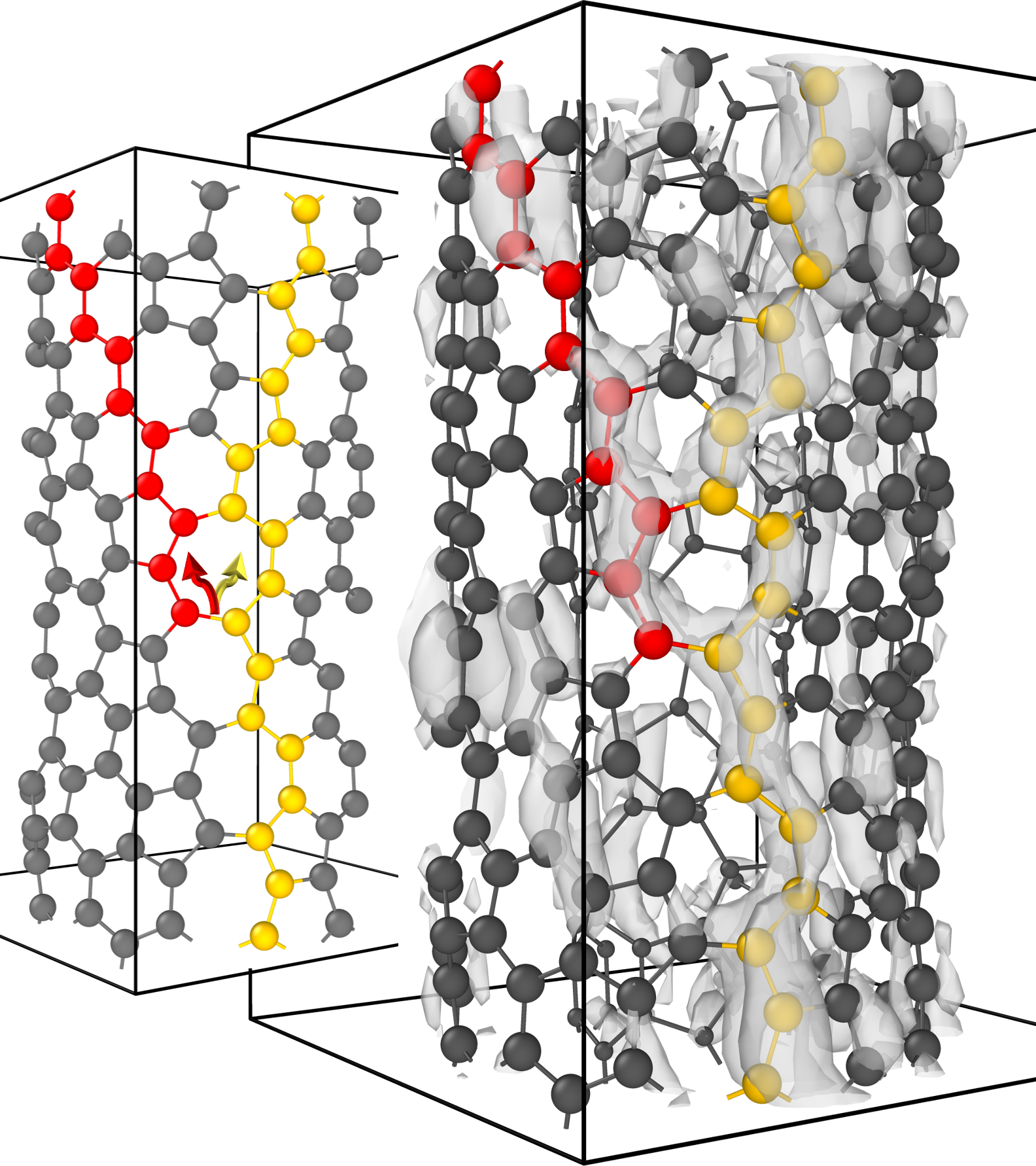}
    	\caption{Spatially projected electronic conductivity in a-CNT. The presence of the 7-member ring defect causes a bifurcation of the electronic conduction path on the yellow-colored C atom into the red-colored atom.}
	\label{fig:Cfig_SPC}
\end{figure}

The electronic density of states (EDoS) for a-CNT was computed within \texttt{VASP} and the extent of localization of Kohn-Sham states ($\phi$) was calculated as the participation ratio (EIPR) using the following equation:

\begin{equation}
    I(\phi_n) =  \frac{\sum_i {| a_n^i |^4  }}{(\sum_i {| a_n^i |^2  } )^2}
    \label{eqn:EDoS}
    \end{equation}

\noindent where a$_n^i$ is the contribution to the eigenvector ( $\phi_n $) from the i$^{th}$ atomic orbital. High (low) values of EIPR indicate localized (extended) states. The Fermi-level ($E_f$) in the EDoS and EIPR plots in Fig. \ref{fig:Cfig_DoS_IPR} was shifted to zero. The EDoS plots for a-CNT do not show any spectral gap at $E_f$, and a few electronic states appear localized around the Fermi level. The non-zero gap (NZG)  observed for a-CNTs has been reported for armchair SWCNTs, as well as SWCNTs with $n - m \neq 3\zeta$, where $\zeta$ is a positive integer \cite{metalicnn,MINTMIRE}. In addition to the NZG similarity in armchair c-CNT and a-CNT EDoS, we found that replicated periodic images of both structures in the $z$-direction share a degree of structural similarity as well (see LEFT] (RIGHT) inset in Fig. \ref{fig:Cfig_DoS_IPR} for a-CNT (c-CNT)). The effect of disorder in the a-CNTs network on the electronic conduction active path was analyzed by projecting the electronic conductivity into a spatial grid. This involved the implementation of the space-projected conductivity (SPC) formalism  which exploits the Kubo-Greenwood formula to obtain information about the conduction pathways in materials \cite{SPC, SPC1}. Fig. \ref{fig:Cfig_SPC} is a representation of the conduction path observed in a-CNT. We found that a-CNT favored only axially directed conduction paths along connected hexagonal carbon rings. The axially directed conduction path is consistent with published results for crystalline nanotubes \cite{dresselhaus1998physical, conductivity1, conductivity2}. While the presence of non-hexagonal rings in a-CNT restricts electron flow in the network, this effect is not as pronounced as it is in amorphous graphite \cite{LAG}. As shown in Fig. \ref{fig:Cfig_SPC}, the 7-member ring does not completely terminate the conduction path but rather acts as a bifurcation of the electron transport on the yellow-colored C atoms into the red-colored C atom path. We stress that the average electronic conductivity of a-CNT decreased by a factor of order 10$^2$ when compared to crystalline CNT \cite{conductivity1,Wang2018}.

\begin{figure}
	\centering
	\includegraphics[width=0.9\linewidth]{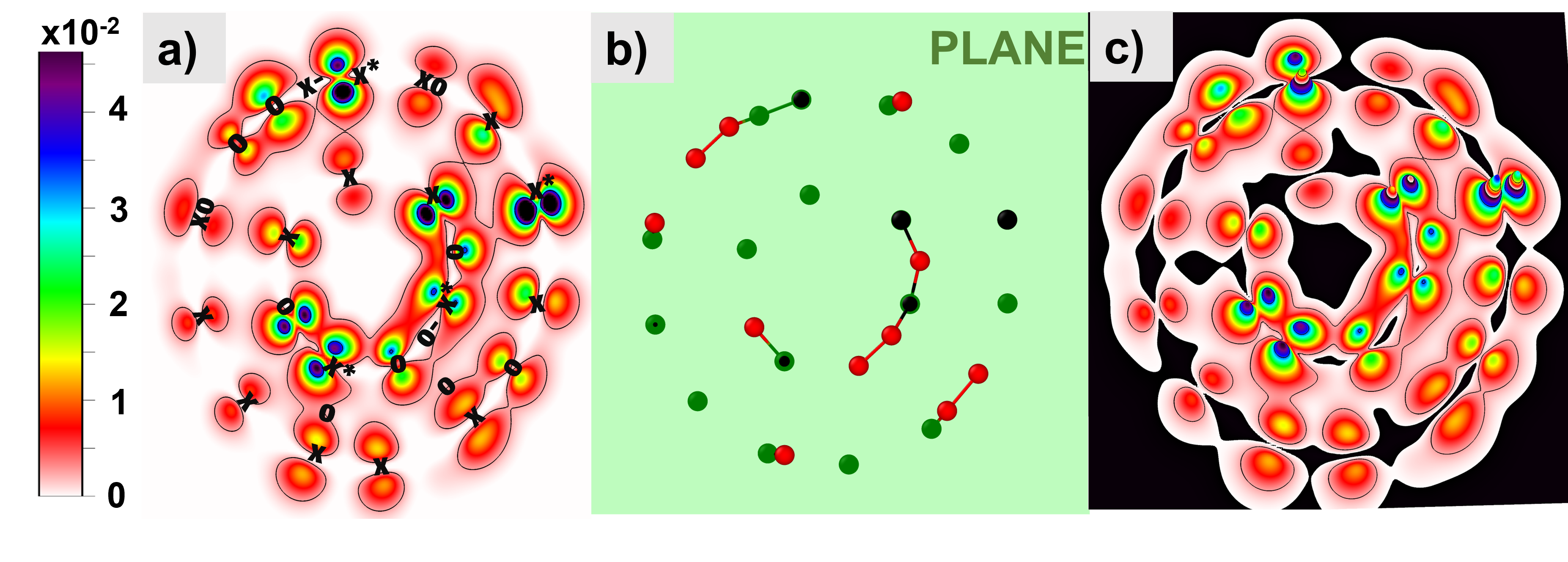}
    	\caption{The figure in (a) shows the $\pi$-electron distribution on a single slice of a 400 atom a-CNT model. The atoms contributing to that particular slice are shown in (b). In (b), the red atoms (R1) are above the green plane (the slice being considered). There are 2 shades of black in (b), the light (B1) and dark (B2) shades correspond to atoms that are below and intersect the green plane respectively. In the plot in (c), the $\pi$-electron density is normalized to the maximum value in the slice and mapped to a mesh grid with zero values (colored as black) to increase the contrast in the  $\pi$-electron density distribution. We have labeled the atoms positions in (a) that corresponds to (b). in (a) \textbf{x}  and \textbf{o} corresponds to B1 and R1 atoms respectively. \textbf{x$^\ast$} are B2 atoms. \textbf{x-} and \textbf{o-} are atoms within the region that do not contribute to the $\pi$ orbital, while \textbf{xo} indicate a combined contribution of two atoms (\textbf{x} and \textbf{o}).  The figure shows a slice was taken at 1.9 \AA~from the origin. An animation, PiBandDensity\_mov.mp4 in the supplementary material \cite{suppl} shows variations in the $\pi$-electron distribution for multiple slices across the a-CNT$_{400}$ model discussed here.}
	\label{fig:Cfig_Piband}
\end{figure}

Inter-layer cohesion in layered carbon structures derives from a combination of Van der Waals forces and delocalized electrons originating from occupied states near the Fermi level ($E < E_f$). These states are the linear combination of $\pi$-orbitals that are projected in the normal direction in the locality of each sp$^2$ carbon atom.  For the a-CNT models, the contribution of the electrons in the $\pi$ subspace was decomposed from the total electronic charge density and projected as isosurfaces on sliced partitions. Fig. \ref{fig:Cfig_Piband} shows the $\pi$-electron distribution for ten bands closest to the Fermi level. Fig. \ref{fig:Cfig_Piband} (a) shows the $\pi$-electron distribution on a single slice of a 400 atom a-CNT model. The atoms contributing to that particular slice are shown in (b). The red atoms (R1) are above the green plane which is the slice being considered. there are 2 shades of black in (b), the lighter (B1) and darker (B2) shades of black correspond to atoms that are below and intersect the green plane respectively. The reader can find a 3D visualization of contributing atoms from Fig S5 \cite{suppl} which is a CW rotation of Fig. \ref{fig:Cfig_Piband} (b).  In Fig. \ref{fig:Cfig_Piband} (c), the $\pi$-electron density is normalized to the maximum value in the slice and mapped to a mesh grid with zero values (colored as black) to increase the contrast in the  $\pi$-electron density distribution. We have labeled the atoms positions in Fig. \ref{fig:Cfig_Piband} (a) that corresponds to Fig. \ref{fig:Cfig_Piband} (b). In Fig. \ref{fig:Cfig_Piband} (a), \textbf{o}, \textbf{x} and \textbf{x$^\ast$}  and  corresponds to R1, B1 and B2 atoms respectively. \textbf{x-} and \textbf{o-} are atoms within the region that do not contribute to the $\pi$ orbital. \textbf{xo} indicate a combined contribution of two atoms (\textbf{x} and \textbf{o}) to the electron distribution.  The figure shows a slice was taken at 1.9 \AA~from the origin and Fig. S6 shows the slicing pattern implemented on the a-CNT$_{400}$ model. The choice of 1.9 \AA~is arbitrary and we have provided an animation (PiBandDensity\_mov.mp4) in the supplementary material \cite{suppl} that shows variations in the $\pi$-electron distribution for multiple slices across the a-CNT$_{400}$ model discussed here. Rozp$\l$och and co-workers \cite{36,rozploch2003new} suggested the cross-interactions related to the $\pi$-electrons as a contributor to the cohesion between layers in the metallic model of Schmidt \cite{schmidt} based on the analysis of the g-factor \cite{rozploch2003new}. The low concentration of quasi-free $\pi$-electrons in the gallery causes a weak metallic bond between the tubes. 

For a-CNT, we found that the maximum value for the $\pi$-electron distribution is $\approx$ 10 \%  of the maximum total charge distribution in all the bands and $\approx$  2.75 \% of the maximum total charge distribution around the center of the gallery. While the low charge densities in the gallery may lead one to argue that the interaction between tubes in CNT originates purely from Van der Waals interaction, the simulation temperature ($\approx$ 3000 K), obtained average electrical conductivity, and (to be discussed) thermal conductivity of our models, coupled with the absence of corrections for Van der Waals interaction in the ML-GAP (or conventional DFT) potential proves otherwise. In the simulation of amorphous graphite \cite{LAG}, the inclusion of Van der Waals corrections in the \texttt{VASP} implementation \cite{DFT_D} did not change the average energy of the large systems obtained using ML-GAP. This support the argument that the "metallic" cross-interaction of the delocalized $\pi$-electrons is involved in the inter-layer cohesion. Additionally, the values observed at the center of the gallery in a-CNTs were lower than those from amorphous multi-shell fullerenes ( $\approx$ $4.3\%$) \cite{BO} but larger than amorphous graphite ($\approx$ $2.0\%$) \cite{LAG}. The mid-gallery values for these amorphous carbon allotropes are considered to be subjected to the influence of the local atomic curvature on the degree of the linear combination (cross-interaction) of the Kohn-Sham orbitals describing the $\pi$-bands. It is noteworthy that in the amorphous phase of these layered carbon structures, layering persists despite the absence of an exact graphite stacking registry.

\subsection{\label{sec:phonon}Vibrations and Thermal conductivity} 

\begin{figure}
	\includegraphics[width=.9\linewidth]{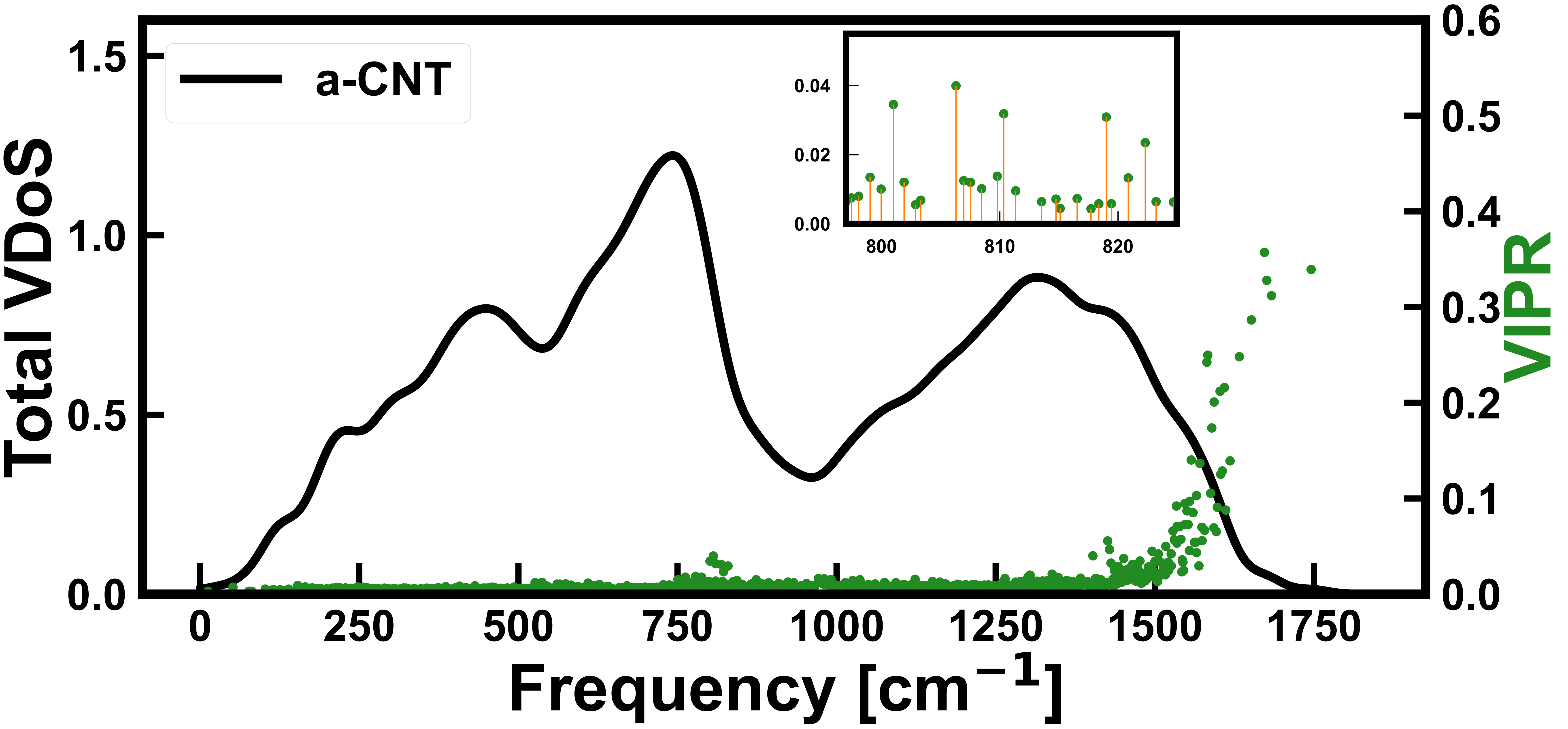}
	\caption{Figure showing the total VDoS and VIPR for a-CNT calculated from the harmonic approximation. The inset shows the VIPR around 797 cm$^{-1}$ 825 cm$^{-1}$  }
	\label{fig:Cfig_VDOS_VIPR}
\end{figure}

\begin{figure*}
	\centering
	\includegraphics[width=0.9\textwidth]{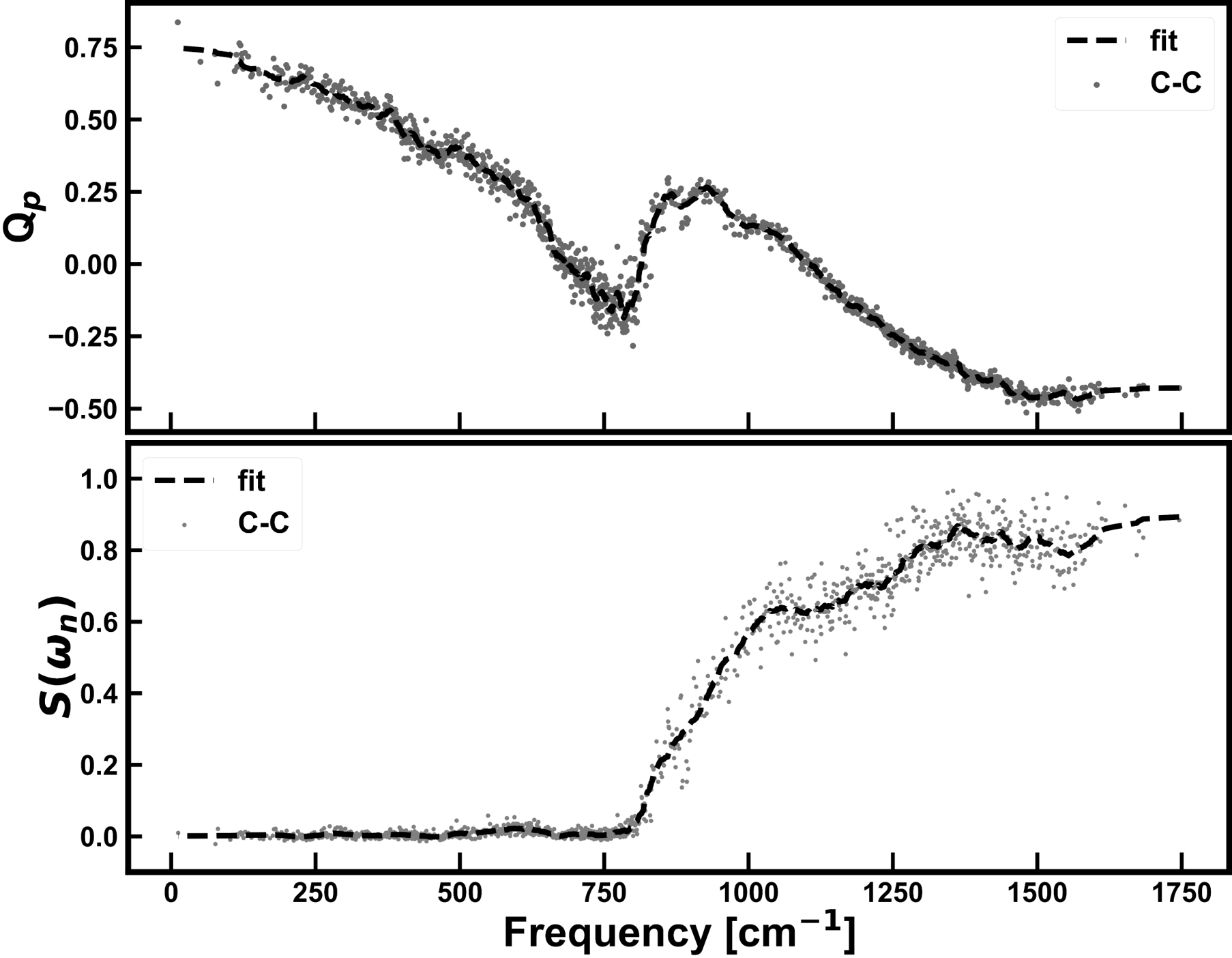}
    	\caption{Figure showing the [TOP] phase quotient and [BOTTOM] stretching character of a-CNT}
	\label{fig:Cfig_PQBS}
\end{figure*}

Carbon nanotube samples that are produced in the laboratory are not always perfect crystals and may include non-hexagonal carbon rings. Additionally, there could be impurities (like fullerenes, graphitic and amorphous carbon, carbon nanoparticles, etc.) in the samples \cite{impurities} and this can lead to confusion in the interpretation of experimental vibrational spectra. We present in Fig \ref{fig:Cfig_VDOS_VIPR}, a vibrational spectrum that is potentially an experimentally testable fingerprint of a-CNT. The atomic vibrational density of states was calculated within the harmonic approximation model as: 

\begin{equation}
    g(\omega) = \frac{1}{3N} \sum_{i=1}^{3N} \delta(\omega - \omega_i) 
\end{equation}

\noindent where, $N$ and $\omega_i$ represent the number of atoms and the eigen-frequencies of normal modes, respectively. The $\delta$ function (approximated by a Gaussian with a standard deviation equal to 1.5\% of the maximum frequency) ensures that high-density values were assigned to vibration frequencies that lie close to the normal modes. The (green) scatter plot in Fig. \ref{fig:Cfig_VDOS_VIPR} is the vibration inverse participation ratio (VIPR) that indicates the extent of localization of each normal mode frequency \cite{BO}. Low values of VIPR indicate vibrational mode evenly distributed among the atoms while higher values imply that few atoms contribute at that particular eigen-frequency. Additionally, the region where VIPR $\gtrsim$ 0.15 is considered the transition frequency from diffusons to locons \cite{Henry35}, and is found to be at around 1582 cm$^{-1}$ in a-CNT.  The lack of periodicity in the a-CNT lattice restricts vibrations to non-propagating modes (e.g. diffusons and locons) \cite{phonons1, phonon2} and does not allow for a rigorous classification of the vibrational modes as purely acoustic (at low-frequency spectrum) or purely optical (at high-frequency spectrum) as in crystals. However, the phase quotient ($Q_p$) of Bell and Hibbins-Butler \cite{Bell_1975} provides a measure of how vibrations of neighboring atoms are in-phase (acoustic mode) and out-of-phase (optical mode). The normalized $Q_p$ is given as \cite{Allen_Feldman}:

\begin{equation}
    Q_p ~= \frac{1}{N_b}\frac{\sum_{m} \mathbf{u}^{i}_{n} \cdot \mathbf{u}^{j}_{n}}{\sum_{m}| \mathbf{u}^{i}_{n} \cdot \mathbf{u}^{j}_{n}|}  
    \label{eq:PQ}
\end{equation}

\noindent where $N_b$ is the number of valance bonds, $\mathbf{u}^{i}_{n}$ and  $\mathbf{u}^{j}_{n}$ are the normalized displacement vectors  for the $n^{th}$ normal mode. The index, i, sums over all the C atoms and $j$ enumerates neighboring atoms of the $i^{th}$ atom. The  in-phase  vibration of the bulk material gives $Q_p$ = 1 (purely acoustic). Conversely, a value of -1 would correspond to motion in the opposite direction between neighboring atoms (purely optical). It then follows that positive (negative) $Q_p$ is more ``acoustic-like” (``optical-like”). Another important  tool for understanding vibrations is the bond-bending or bond-stretching character ($S(\omega)$) given as: 

\begin{equation}
    S(\omega_n) ~= \frac{\sum_{m} |\mathbf{u}^{i}_{n} - \mathbf{u}^{j}_{n}| \times \hat{\mathbf{r}}_{ij}}{\sum_{m}| \mathbf{u}^{i}_{n} - \mathbf{u}^{j}_{n}|}  
\end{equation}

\noindent where, $\mathbf{u}^{i}_{n}$ is as the same as in Eq. \ref{eq:PQ}and $\hat{\mathbf{r}}_{ij}$ is the unit vector parallel to the m$^{th}$ bond.  $S(\omega_n)$ is close to unity when the mode of vibration is predominantly of bond-stretching type and will be close to 0 otherwise. Fig. \ref{fig:Cfig_PQBS} show $Q_p$ [TOP] and $S(\omega_n)$ [BOTTOM] for a double-walled a-CNT (DWCNT) with 400 atoms. The relationship becomes that optic-like modes at the high-frequency end have bond-stretching character and acoustic-like modes at the low-frequency end have bond-bending character. We have provided animations to visualize the vibrations in the supplementary material. The mid-frequency region around 800 - 820 cm$^{-1}$ show some localization (see inset in Fig. \ref{fig:Cfig_VDOS_VIPR}) and a local minimum at $\approx$ 807 cm$^{-1}$, which is also $\approx$ 0 in the for $Q_p$ plot. It is worth mentioning that the behavior of $Q_p$ around 0 cannot be distinguished into acoustic- or optical-like modes. The animation, freq\_807.mp4, shows that the modes in this region are quasi-localized ``resonant modes" \cite{Feldman_Allen} since they are distributed over a number of atoms and also are not diffuson to locon transition frequencies. Other notable vibrations in the double-wall a-CNT analyzed are at 32, 44, 55 cm$^{-1}$ which correspond to twisting vibrations of the outer shell, translation, and twisting vibration of the inner shell respectively (see supplementary material \cite{suppl} for animations of these modes). We emphasize that the ``twisting" modes of a  particular tube at a given frequency cost very little energy. The electron gas in the "gallery" between layers is quite diffuse and it is weakly correlated with particular atoms in the layer, so the energy dependence of the system, to the first approximation, is on the distance between the shells and not affected by the angle. The radial breathing modes (RBMs) which dominate the low-frequency Raman spectrum in experiments are unique to specific MWCNTs and much focus has been given to the research of RBM because of its importance in the applications of structure and property characterizations \cite{dresselhaus2005raman,RBM_lowFreq,RBM_lowFreq_contradict}. We observed an in-phase, low-frequency RBM at around 81 cm$^{-1}$ (see animation in supplementary material). RBM in MWCNTs has been observed experimentally from Raman-active peaks around 100 cm$^{-1}$ \cite{RBM_lowFreq}, and in-phase RBM at around 94 (85) cm$^{-1}$  have been predicted for armchair DWCNT (SWCNT) \cite{RBM_lowFreq_simulation}. We note here that the similarities observed between a-CNT and the armchair CNT could be a path worth exploring. 

Finally, we shift focus to the contribution of the topological defect to the thermal conductivity (TC) in a-CNTs. Heat conduction in carbon materials is usually dominated by phonons (even in graphite \cite{usePhonons}), this allows us to consider the contribution of the heat flux (\textbf{J}) for each atom \cite{heatflux1, heatflux2} in a 2000 atom a-MWCNT and then relate an ensemble average of the auto-correlation of \textbf{J} to the TC ($\kappa$) using the Green-Kubo formala given as \cite{Green, Kubo}:

\begin{equation}
    \kappa ~= \frac{1}{3Vk_BT^2}\int_{0}^{\tau} \left < \textbf{J} (0) \cdot \textbf{J} (t) \right > dt
    \label{eq:kappa}
\end{equation}

where V, T, and $k_B$ are the system volume, temperature, and Boltzmann's constant, respectively. The upper limit of the integral was approximated by $\tau$ (= 2 ns) which is the correlation time required for the heat current auto-correlation to decay to zero. The TC was obtained by averaging the integral in Eq. \ref{eq:kappa} from 10 independent models. The Nos\'e–Hoover thermostat \cite{nose, hoover} was used for thermalization and equilibration at T = 300 K at a fixed volume using a 1 fs time-step, and at the beginning of the simulation, initial velocities were assigned to the atoms randomly from a Gaussian distribution. Our result showed that the average TC calculated for a-CNT was 22.15 Wm$^{-1}$K$^{-1}$ at 300 K. The thermal conductivity of MWCNT is $\approx$ 3000 Wm$^{-1}$K$^{-1}$ \cite{MWCNT_therma1_3K, MWCNT_therma2_3K}, so the thermal conductivity of a-MWCNT is an order of 10$^2$ less than c-MWCNT, which is similar to what we have reported for the electrical conductivity.

\section{Conclusion}
In this work, we explored the formation of a-CNT starting from an initially random configuration of carbon atoms using the ML-GAP potential. The starting configuration required terminating the periodic boundary conditions in the $x$-$y$~plane only. The complexity involved in the choice of the important variables to simulate realist a-CNTs was minimized by learning the important features from a random forest implementation. The structure of the a-CNT models was validated by conjugate gradient relaxation as implemented within SIESTA and VASP. The diameter of the a-CNTs ranged from 0.5 nm in a-SWCNTs to 2 nm in a-MWCNTs, and the inter-wall spacing was $\approx$ 0.31 nm. The ring defects in a-CNTs were discussed and a potentially new type of defect was observed. Electronic structure analysis showed that there was no band-gap at the Fermi level, which is a property observed in all armchair CNTs. The delocalized $\pi$-electrons in the gallery were confirmed to be involved in the inter-layer cohesion in a-CNTs. The density of state and corresponding participation ratio for the phonon vibrations were analyzed, and the result showed that a-CNTs have localized states only at the high-frequency end of the vibration spectrum which is consistent with other amorphous structures. The phase quotient and stretching character analysis further suggested that those localized sites were from atoms participating in non-hexagonal rings. The average thermal conductivity for a-CNT was 22.15 Wcm$^{-1}$K$^{-1}$ at room temperature and is an order of 10$^2$  less than crystalline MWCNT, which was also similar to the average thermal conductivity.  In conclusion, the strong proclivity to form layered structures, even with topological defects,  from chaotic configuration remains one of the wonders of carbon. 

\medskip
\textbf{Supporting Information} \par 
Supporting Information is available in the supplementary material \cite{suppl} for this manuscript.

\medskip
\textbf{Acknowledgements} \par 
The authors thank Anna-Theresa Kirchtag for proofreading the manuscript. The U.S. Department of Energy for support under Grant No. DE-FE0031981 and XSEDE (supported by National Science Foundation Grant No. ACI-1548562) for computational support under allocation no. DMR-190008P

\medskip

%
\bibliographystyle{MSP}
\bibliography{aCNT}

\newpage


\begin{table}
	\begin{center}
		\caption{Table showing the important Feature index predicted from a random forest classifier with K=4 cross-validation set. It also shows the accuracy rate on the test set for each fold and the mean and standard deviation calculated for scaling the aspect ratio and system size for the decision boundary construction. See discussion in the supplementary material \cite{suppl}}\label{tab:Ctable_RF}
		\begin{tabular*}{\linewidth}{@{\extracolsep\fill}ccccccc@{\extracolsep\fill}}
			\hline
			 & \textbf{Kfold-1} & \textbf{Kfold-2} & \textbf{Kfold-3} & \textbf{Kfold-4} & \textbf{mean} & \textbf{std}\\
			\hline
   
			\textbf{Aspect Ratio (height / diameter)} & 0.32	& 0.34 &	0.32 &	0.33 & 4.3 & 1.25 \\
			
			\textbf{System Size (No. of atoms)} & 0.18 &	0.20 &	0.19 &	0.23   & 1672 & 883 \\
			
			\textbf{Cutoff Radius [\AA]} & 0.13 &	0.12 &	0.12 &	0.08   &  &  \\
			
			\textbf{Vacuum in the $x$-$y$~plane [\AA]} & 0.11 &	0.12 &	0.12 &	0.13 &   &   \\

			\textbf{Energy [eV/atom]} & 0.25 &	0.22	& 0.25 &	0.23 &  &  \\
            \hline
               \textbf{Accuracy [\%]} & 87 &	92	& 87 &	89 &  &  \\
            \hline
		\end{tabular*}
	\end{center}
\end{table}

\end{document}


\pagestyle{fancy}
\rhead{\includegraphics[width=2.5cm]{vch-logo.png}}

\title{Supplementary Material: Formation of amorphous carbon multi-walled nanotubes from random initial configurations}

\maketitle


\author{Chinonso Ugwumadu*,}
\author{Rajendra Thapa,}
\author{David A. Drabold}

\begin{affiliations}
C. Ugwumadu, Prof. D. A. Drabold\\
Department of Physics and Astronomy, \\
Nanoscale and Quantum Phenomena Institute (NQPI),\\
Ohio University, Athens, Ohio 45701, USA\\
E-mail: cu884120@ohio.edu\\

Dr. R. Thapa\\
Institute for Functional Materials and Devices\\
Lehigh University,Bethlehem, PA 18015, USA
\end{affiliations}


\keywords{amorphous solids, nanotubes, carbon}

\section{Description of Animations produced for the amorphous Nanotubes (a-CNT) models}

We have produced some animations to aid the reader in visualizing some of the discussions in the paper.  The animations can be found \href{https://people.ohio.edu/drabold/nanotubes/}{here} or by visiting the url: \url{https://people.ohio.edu/drabold/nanotubes/}

The descriptions are as follows:

\begin{enumerate}

    \item \textbf{README.txt}: File containing similar descriptions of the animations written here. 
    \item \textbf{Hollow\_a-CNT\_mov.mp4}: This describes the formation process of hollow a-CNT.
    \item \textbf{Capped\_a-CNT\_mov.mp4}: This describes the formation process of capped a-CNT. 
    \item \textbf{PiBandDensity\_mov.mp4}: This shows the variations in the $\pi$ electron distribution along the z-direction of  an a-CNT$_400$ model, the slices are perpendicular to the z-direction as shown in the image (a-CNT\_400atoms.jpg) attached to the animation.
    \item \textbf{freq\_vibrations.mp4}: This shows the evolution of all vibrational modes in a-CNT, from high frequency (in-plane) to low frequency (out-of-plane) to Goldston modes (collective vibrations).
    \item \textbf{freq\_11.mp4}: One of the Goldstone modes. This helps to compare other low-frequency modes to a typical Goldstone mode
    \item \textbf{freq\_32.mp4}:  Twisting vibrational frequency of the outer tube, the inner tube does not move 
    \item \textbf{freq\_44.mp4}:  Translational vibration of inner tube
    \item \textbf{freq\_55.mp4}: Twisting vibrational frequency of the inner tube, the outer tube does not move
    \item \textbf{freq\_81.mp4}:  Collective low-frequency radial breathing mode (RBM) od both tubes in the a-DWCNT
    \item \textbf{freq\_807.mp4}: Resonant mode (quasi-localized modes that diffuse away) since the VIPR < 0.15 at this frequency.
    \item \textbf{freq\_1746.mp4}: Locons in high frequency optic-like mode. The vibrating atoms are in red and the direction of vibration is specified using arrows 
\end{enumerate}


\begin{figure}
  \includegraphics[width=0.6\linewidth]{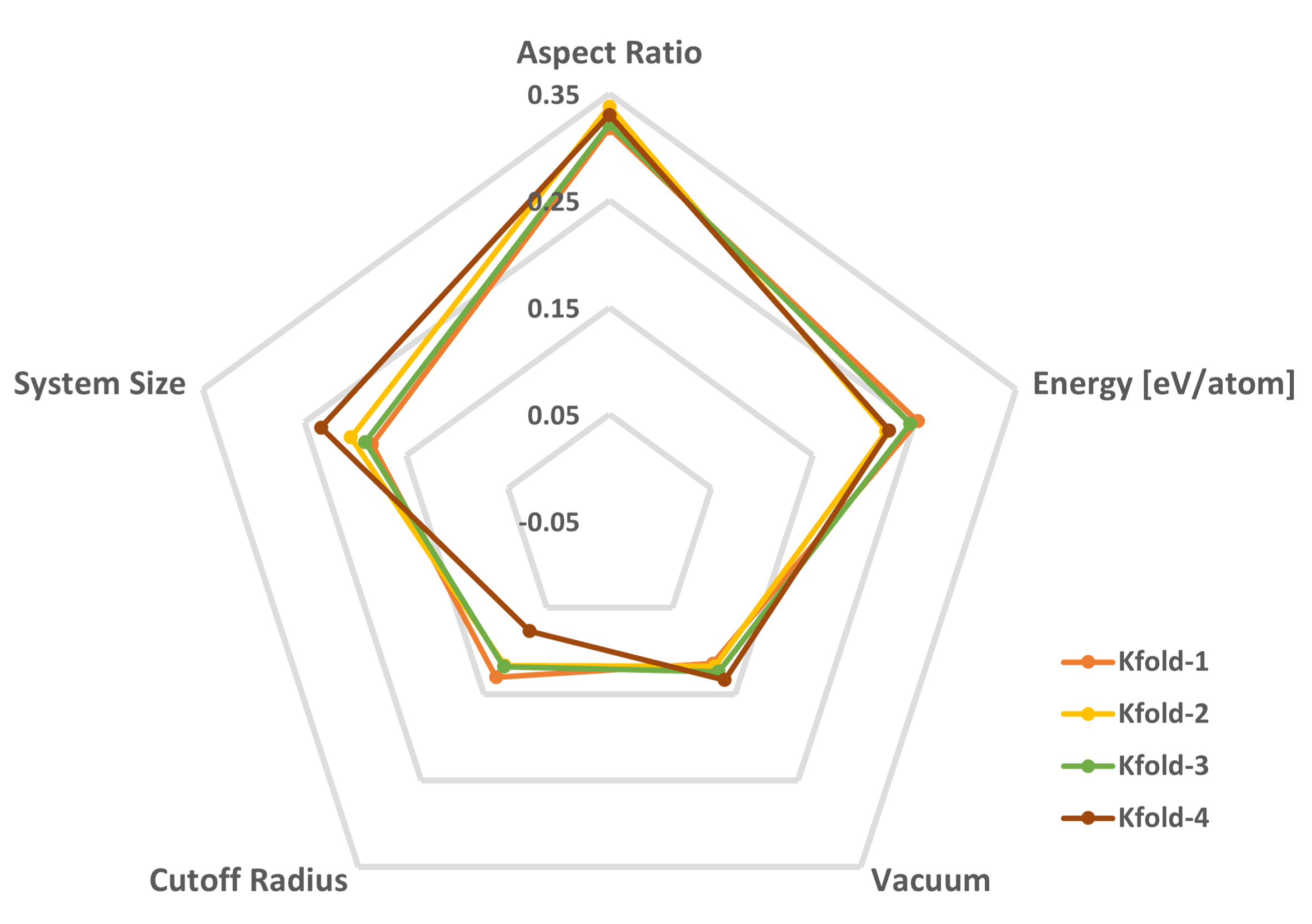}
  \caption{Radar Chart showing the important features from K-fold cross-validation (K =4) of a random forest classifier using the Gini index.}
  \label{fig:CSfig_FeaturesRF}
\end{figure}

\begin{figure}
  \includegraphics[width=.8\linewidth]{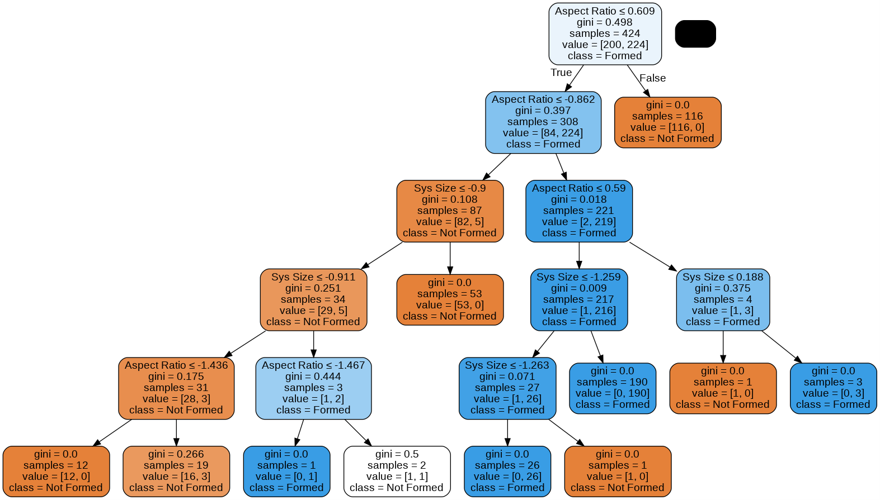}
  \caption{visualization of the decision tree for a binary classifier using the aspect ratio and system size. "Formed" class indicates that a-CNT is formed and the  "Not Formed" class indicates otherwise. see Fig. \ref{fig:CSfig_examples} for examples of such models.}
  \label{fig:CSfig_Tree}
\end{figure}

\begin{figure}
  \includegraphics[width=.8\linewidth]{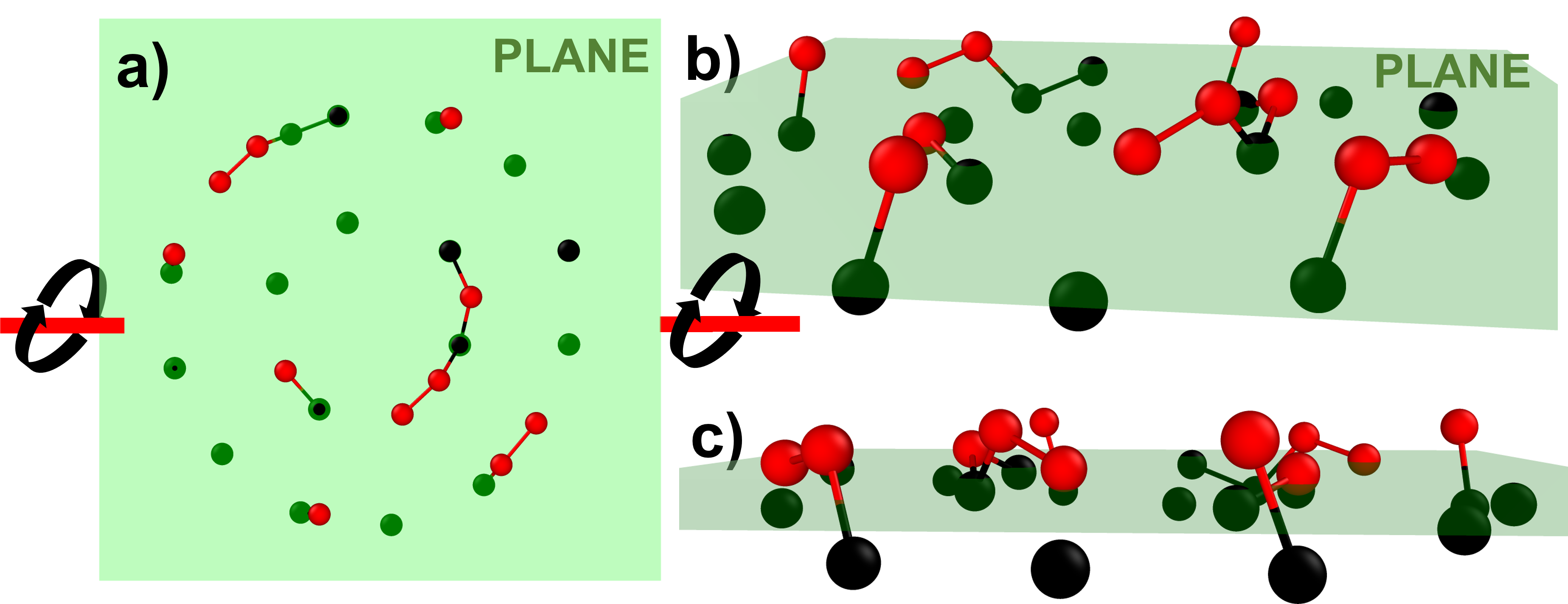}
  \caption{Visualization of the atoms contributing to the $\pi$ electron cloud in the slice discussed in the manuscript. In (a) the black (gray) atoms are below (above) the slice plane. The CW rotation of (a), along the horizontal axis of the plane (shown in red solid lines), is shown in (b) (45$^\circ$ CW) and (c) (90$^\circ$ CW). Atoms that fall right on the plane (see c) are also colored in black.}
  \label{fig:CSfig_atomsContributing}
\end{figure}

\begin{figure}
	\centering
	\includegraphics[width=0.5\linewidth]{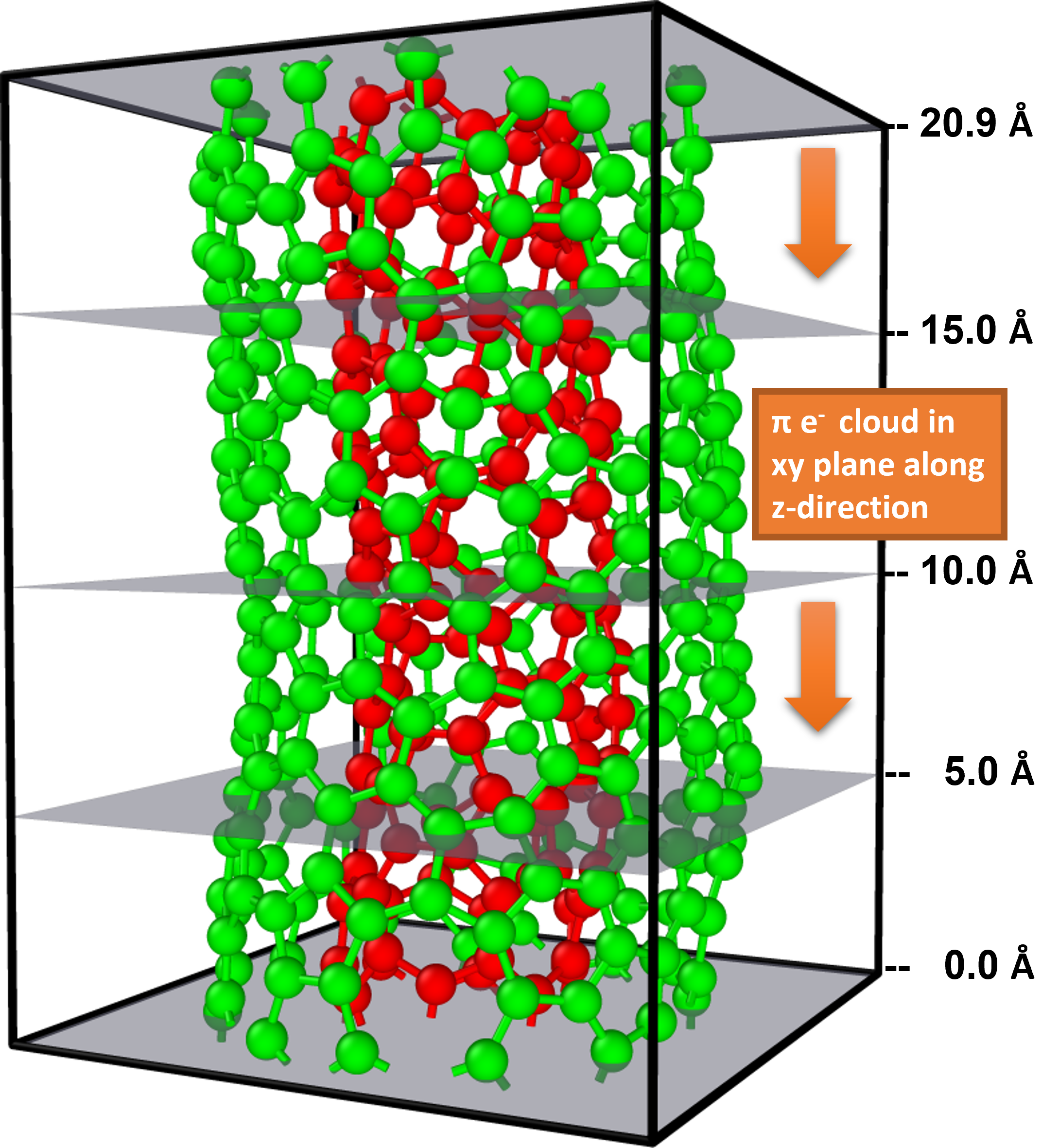}
    	\caption{Figure showing the direction of slicing for one of the a-CNT$_{400}$ models used in the analysis of the $\pi$ electron density distribution.}
	\label{fig:CSfig_nnSlicing}
\end{figure}